# Is the United States losing ground in science?

# A global perspective on the world science system

Loet Leydesdorff [a] and Caroline Wagner[b]

**Abstract**

Based on the *Science Citation Index–Expanded* web-version, the USA is still by far the strongest nation in terms of scientific performance. Its relative decline in percentage share of publications is largely due to the emergence of China and other Asian nations. In 2006, China has become the second largest nation in terms of the number of publications within this database. In terms of citations, the competitive advantage of the American "domestic market" is diminished, while the European Union (EU) is profiting more from the enlargement of the database over time than the US. However, the USA is still outperforming all other countries in terms of highly cited papers and citation/publication ratios, and it is more successful than the EU in coordinating its research efforts in strategic priority areas like nanotechnology. In this field, the People's Republic of China (PRC) has become second largest nation in both numbers of papers published and citations behind the USA.

[a] Amsterdam School of Communications Research (ASCoR), University of Amsterdam, Kloveniersburgwal 48, 1012 CX Amsterdam, The Netherlands; loet@leydesdorff.net; http://www.leydesdorff.net.

[b] SRI International, 1100 Wilson Boulevard, Arlington, VA, 22209, USA, and George Washington University, cswagner@gwu.edu .

## 1. Introduction

The last decade has witnessed significant changes in locations where science is conducted. Data show exponential growth in the share of the People's Republic of China (PRC) for almost all science and technology indicators (Jin & Rousseau, 2004; Zhou & Leydesdorff, 2006). Kostoff (2004) argued that a number of indicators show the PRC outperforming the USA in 2004 in strategic areas like nanotechnology. Using a number of these indicators, Shelton & Holdridge (2004) issued a warning that the USA is losing to competition from an expanding European Union. In the meantime, serious concern has been voiced about whether the USA is declining not only in terms of relative shares (because of the zero-sum game) but also in terms of absolute numbers, for example, on the crucial indicator of performance in terms of scientific publications (Grens, 2006; NSB, 2006; Shelton, 2006).

In this study, we present recent data that are relevant to this debate. By placing the data for recent years (including 2006) in the context of developments over the last decade (King, 2004), we are able to show that the USA is still by far the leading nation in the world of science. The numeric lead of the EU-25, which is larger in size and population than the USA, cannot hide the endogenous problems of the EU science system. The rise of the PRC and other smaller Asian nations seems to continue almost predictably along exponential and linear curves, respectively, expanding scientific output rather than pushing out other performers.

## 2. Methods and data

Data are drawn from the expanded version of the *Science Citation Index* available on the Internet as part of the "Web of Science" (http://portal.isiknowledge.com). The publication counts are based on the three scientific document types which can be cited: research articles, reviews, and letters. We used the field tag "cu=" for countries and the delineations of years as provided by the database. Percentages of world share are based on attributing one full point to each country in the case of internationally co-authored papers. For this reason, world shares may add up to more than 100% (Braun *et al*., 1991; Martin, 1991). The EU-25 series is corrected for "within-Europe" co-authorship.

The subset of nanoscience and -technology journals is based on a detailed study of the journal structure in this area (Leydesdorff & Zhou, 2007). Using "betweenness centrality" as a measure of interdisciplinarity (Freeman, 1977, 1978/9; Leydesdorff, forthcoming), ten journals were distinguished as specifically focused on nanoscience and –technology among 38 journals in chemistry and applied physics that are cited by articles in this core group (Table 1). (The analysis is based on the aggregated journal-journal citation data contained in the *Journal Citation Reports* of the *Science Citation Index* 2004.) We use this set of ten journals as a representation of the development of nanoscience and technology for comparative reasons, and add citation data. However, the specialty is not yet stabilized in a core set of journals; papers can be published in a large number of journals in different specialties (Braun *et al*., 2007).

| |
|---|
| *Advanced Materials* |
| *Chemical Physics Letters* |
| *Chemistry of Materials* |
| *Fullerenes, Nanotubes, and Carbon Nanostructures.* |
| *Journal of Materials Chemistry.* |
| *Journal of Nanoparticle Research* |
| *Journal of Nanoscience And Nanotechnology* |
| *Journal of Physical Chemistry B* |
| *Nano Letters* |
| *Nanotechnology* |

**Table 1**. Ten journals as representing nanoscience and nanotechnology as a specialty, on the basis of the *Journal Citation Reports 2004.*

## 3. Results

Figure 1 shows the percentage of world shares of scientific publications for the six major countries and the EU-25 contributing to science.

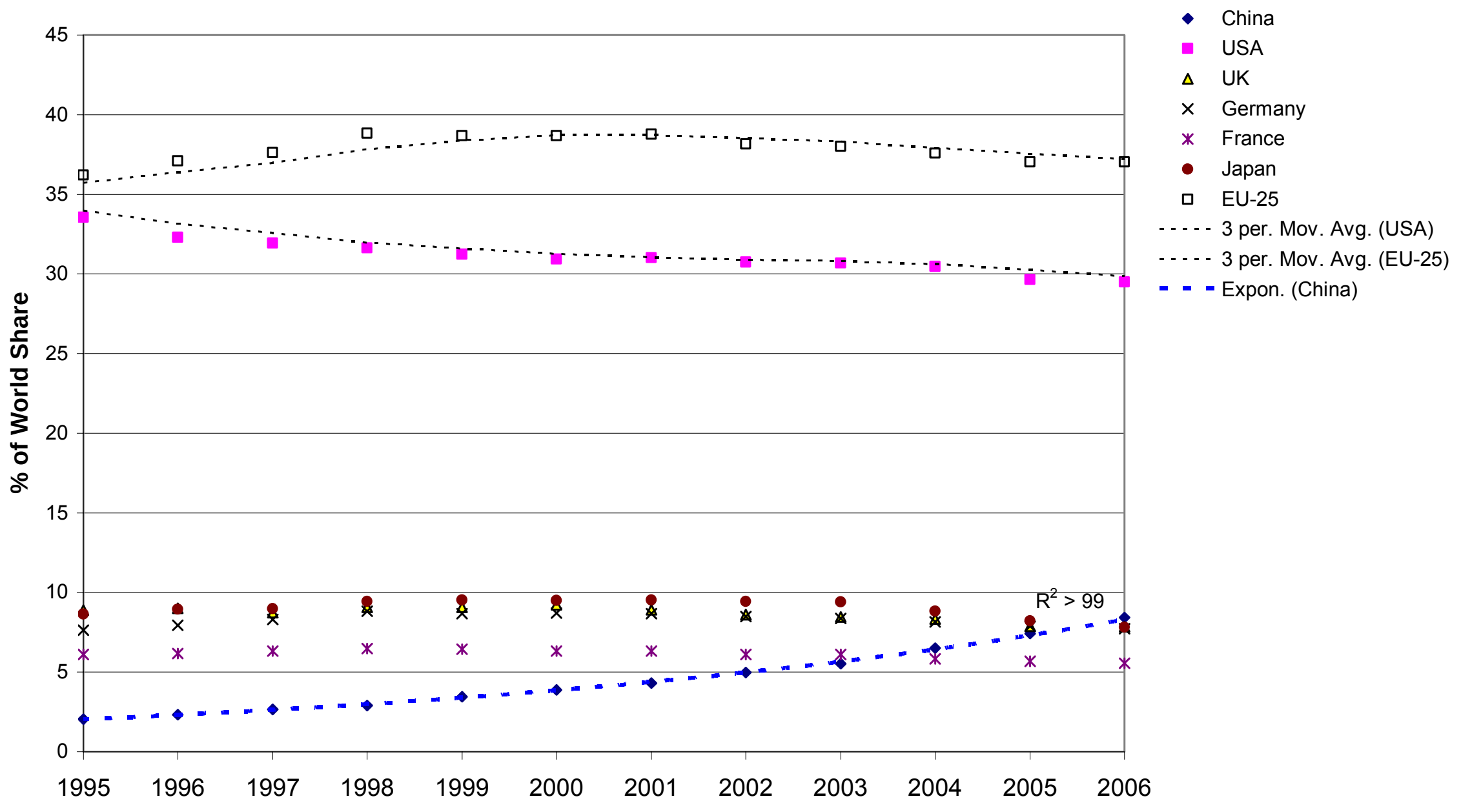

**Figure 1**: Percentages of world shares of publications held by six leading countries and the EU-25, 1995-2006.

The EU expanded from fifteen member states to twenty-five member-states in May 2004,[1] thus the data are reconstructed from the perspective of hindsight. However, Zhou & Leydesdorff (2006) provided data for both the EU-15 and EU-25 until 2004. They conclude that the extension has changed the size of the EU, but not the trends in scientific productivity. Our data show that all the established countries are declining in relative shares. This decline is largely due to the spectacular increase of the percentage share held by China that continues to follow an exponential curve ($r^2 > 0.99$). Using this database, China has a world share of 8.4% in 2006, followed by Japan and the UK with each 7.8%, and Germany with 7.7%. The three-year moving averages added to the lines for both the EU and USA show a similar pattern since 2000.

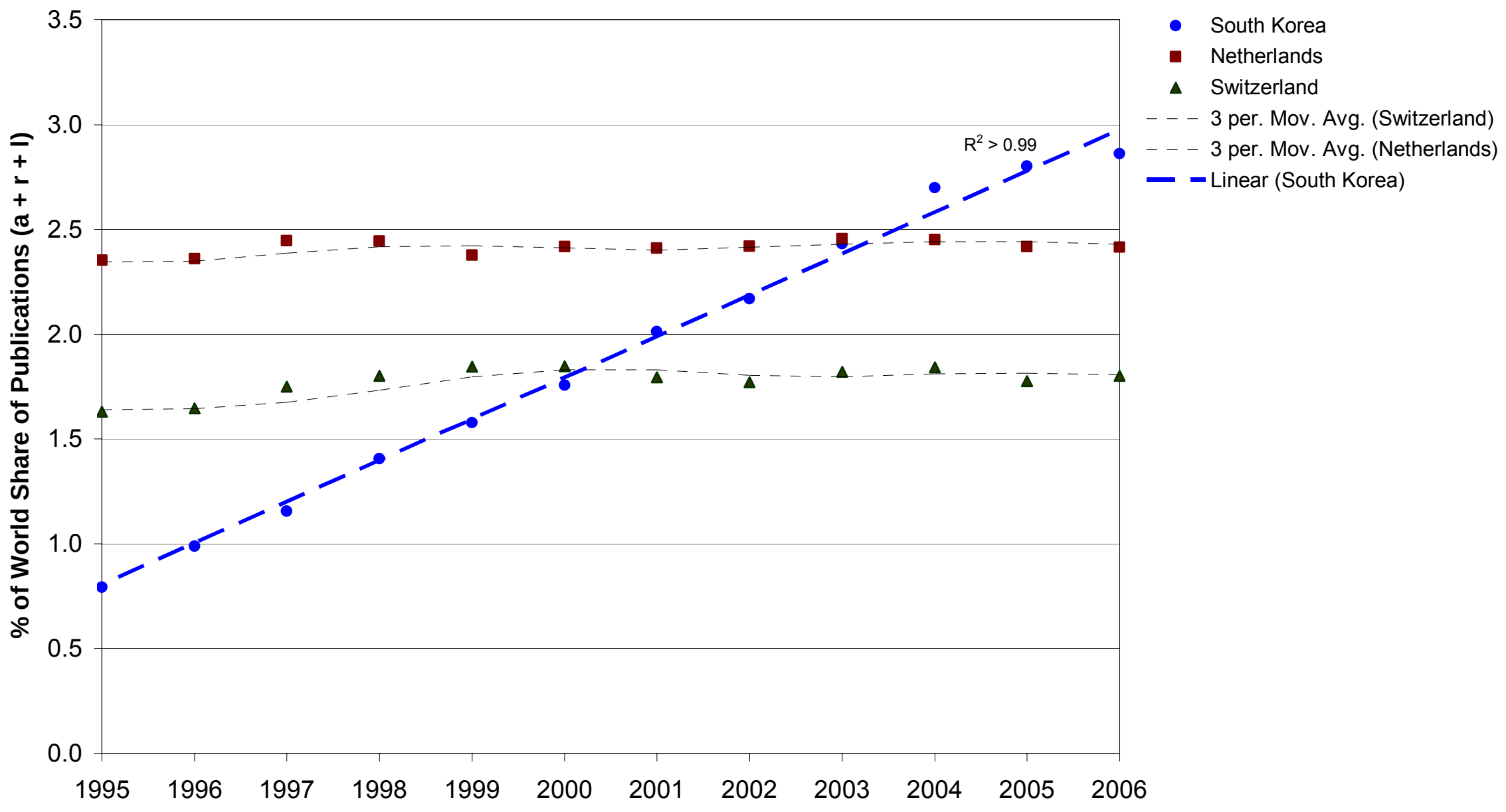

**Figure 2**: Percentages of world shares of publications held by smaller countries, 1995-2005.

[1] Bulgaria and Romania joined the EU on January 1, 2007.

Figure 2 extends the analysis to three smaller countries as examples. Unlike the larger countries, nations like Switzerland and the Netherlands have been able to stabilize their shares. This means that they keep pace with the increased competition, but are no longer able to improve their marginal return (as they did previously). South Korea, however, shows a steady increase. South Korea shares this linear pattern ($r^2 > 0.99$) with other "Asian Tigers" like Taiwan and Singapore (Zhou & Leydesdorff, 2006). In the most recent years, the linear growth of the Korean share seems to level off, but this is not yet significant as a trend breach.

Note that China showed exponential growth (in Figure 1). This spectacular and hitherto sustained pattern of growth may be due to the increasing availability of human capital at Chinese universities and research institutions for publishing in ISI-listed journals, as well as to incentives within China to publish in refereed journals (Zhou & Leydesdorff, 2007).

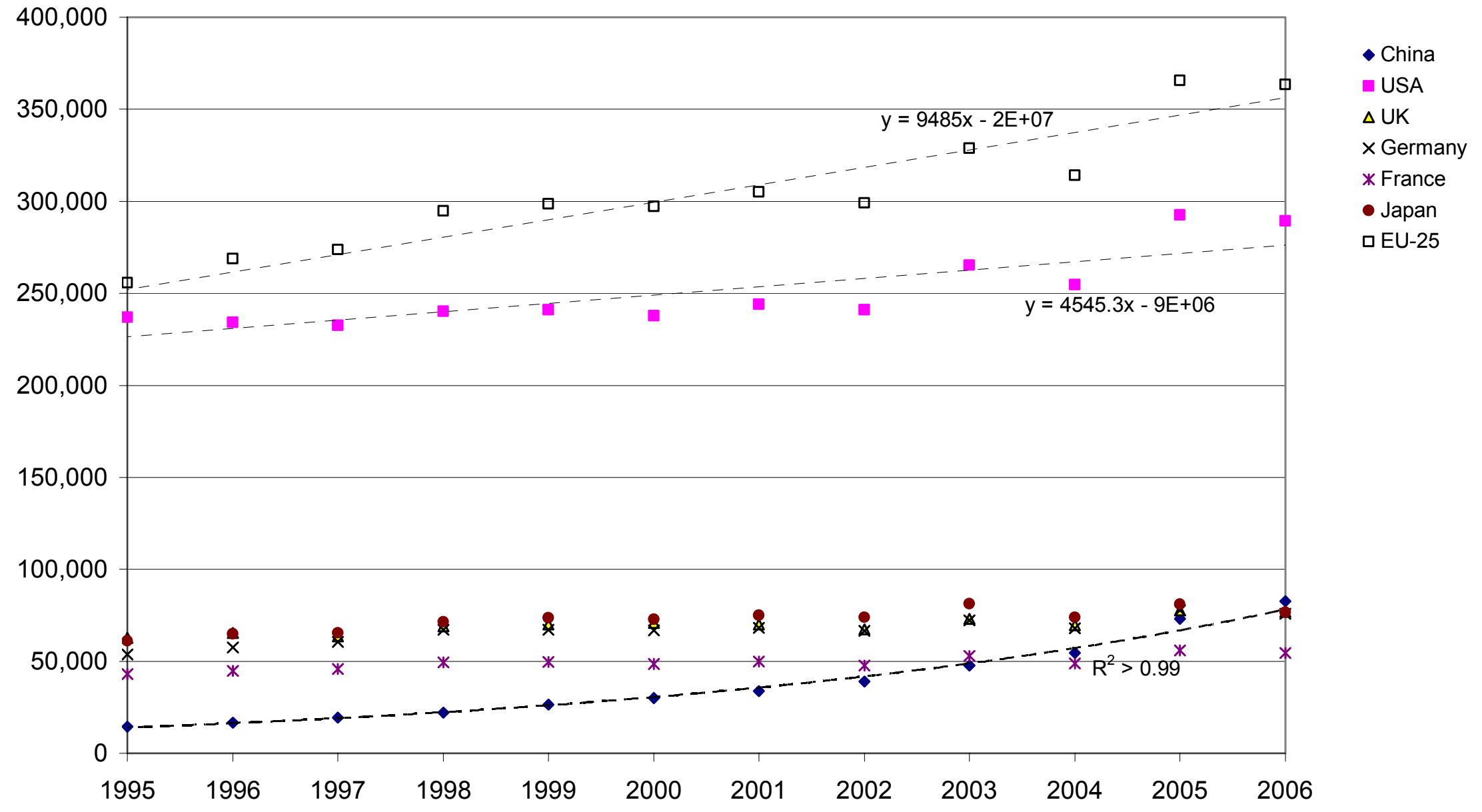

**Figure 3**: Number of publications held by leading countries and the EU-25, 1995-2006.

Figure 3 shows absolute numbers of publications for the six leading countries and the EU-25. The size of the database itself varies from year to year, and this is reflected in the relative shares of major countries. The Chinese contribution again shows an exponential pattern in absolute numbers. The ten-year average of the increase for the EU-25 is more than twice as high as for the USA. The database itself has been expanding by an average of more than 25,000 citable items per year; the USA participates in this increase with approximately 20% and the EU with 41%, but this varies among years (Figure 4).

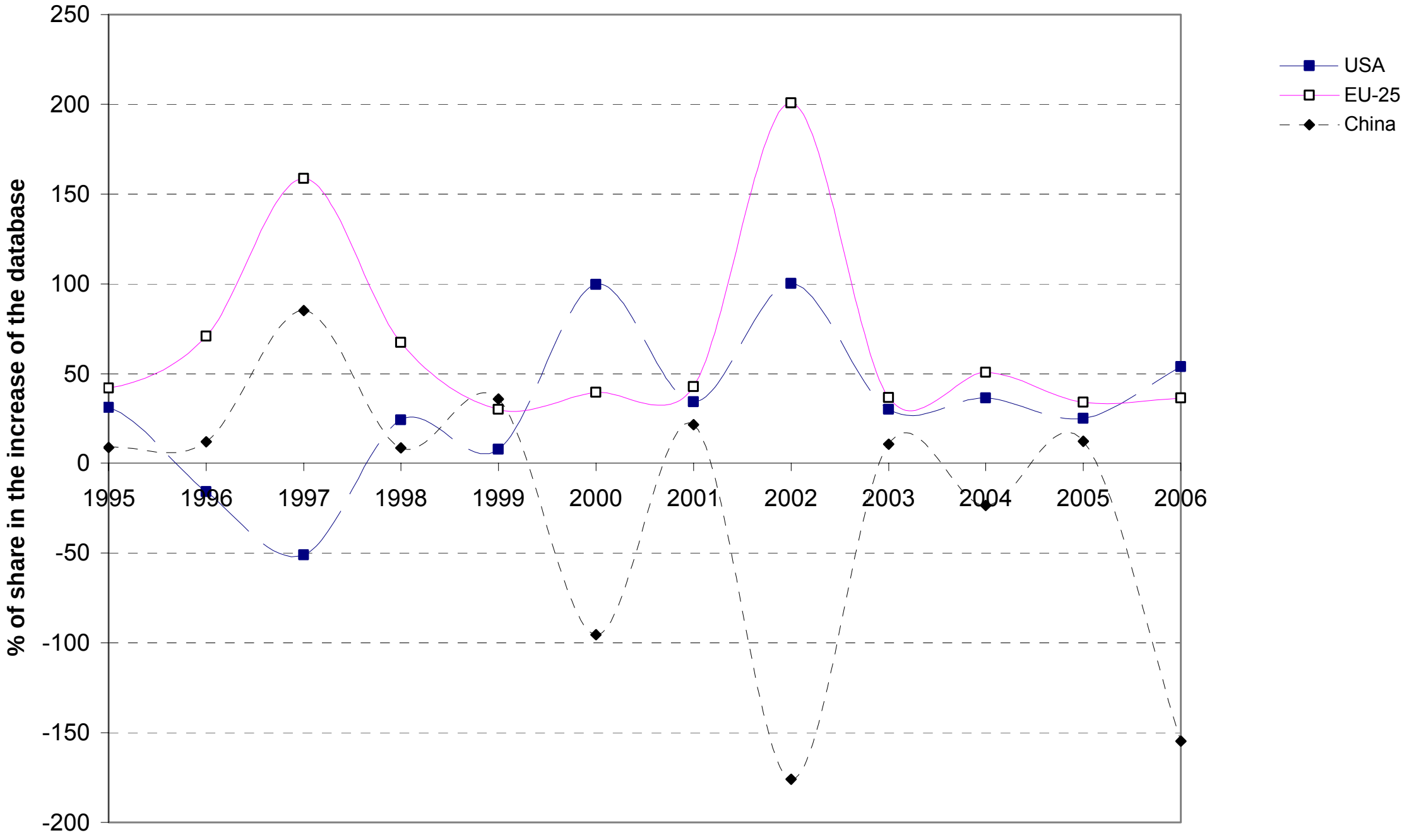

**Figure 4**: Percentage share of publications with an address in the EU-25, the USA, and the PRC in the net increases of the *Science Citation Index–Expanded*, 1995-2005.

Figure 4 shows the participation of the EU-25, the USA, and the PRC in the expansion of the number of publications in the database as a relative percentage, for each two consecutive years. In almost all years, the EU-25 has improved its share of publications more than the USA. The pattern for China is more erratic, showing that mechanisms other than participation in the increase of the size of the database are driving China's performance figures. In other words, China does not profit from the extension of the size of the database, while the EU does.

It is not easy to standardize the measurement of citations using the database online, because citations accumulate with time and the results are therefore dependent on the date of the data collection.  However, on the basis of extensive research (Evidence, 2003), King (2004, at p. 312) provided a table of the 1% most highly cited papers in two periods, which allows us to compare the various nations at these two periods of times, that is, 1997-2001 and 1993-1997 (Figure 5).

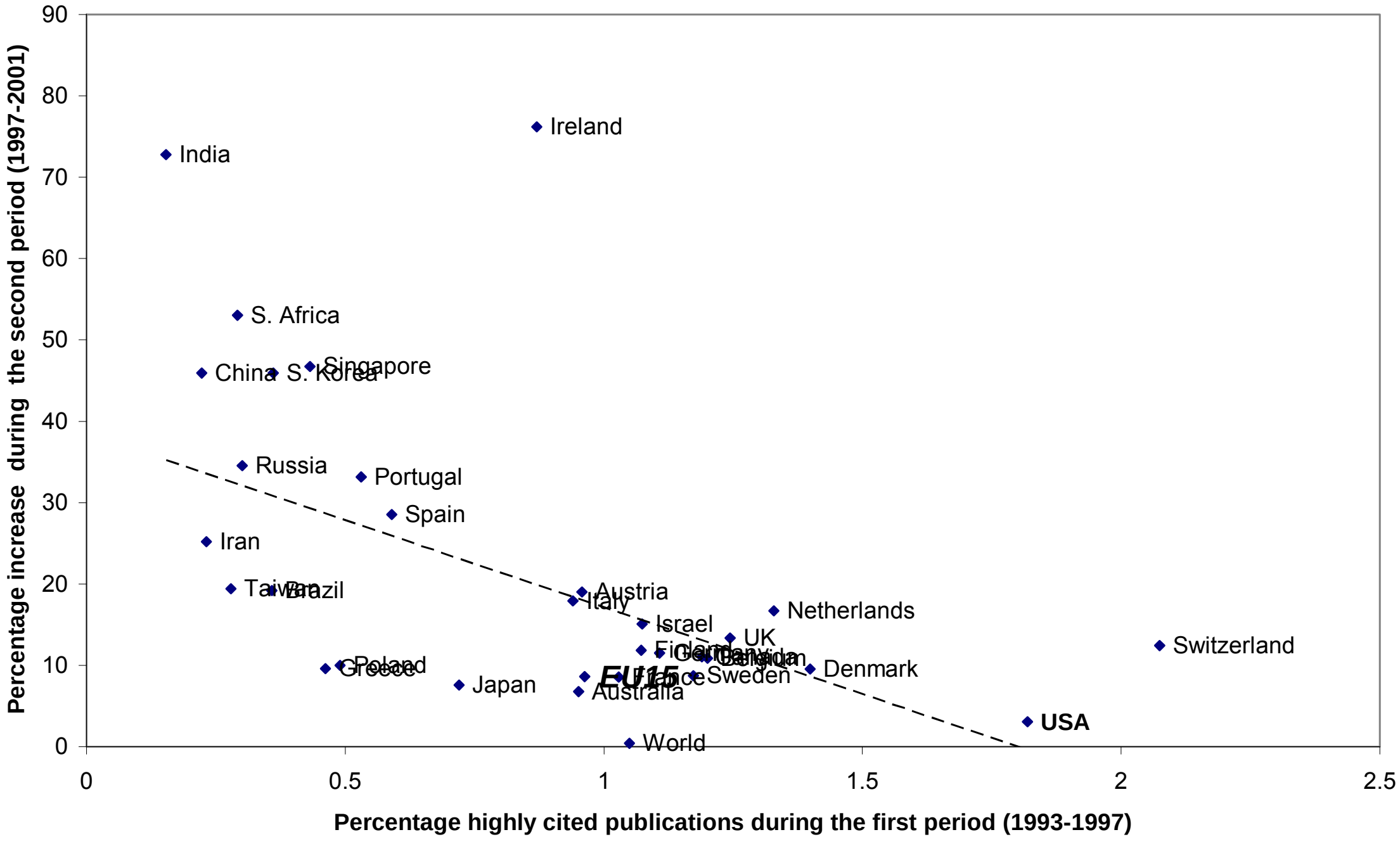

**Figure 5:** Percentage increase of most highly cited publications based on King (2004, at p. 312). Source: Leydesdorff (2005, at p. 412; linear regression line added for the orientation of the reader, *n.s.*).

The USA ranks second largest on this indicator during both periods (behind Switzerland), and like Switzerland it is nevertheless still able to improve its performance. The EU-15 is available in this data and it is highlighted on the map hovering close to Australia. In summary, the USA has been outperforming the EU on this quality indicator by a factor of almost two.

The *Science and Engineering Indicators 2006* of the National Science Board of the USA (NSB, 2006) are based on an analytical version of the ISI-data which has been maintained since 1988 by ipIQ, Inc. (formerly CHI Research, Inc.). In 2003 for example, this data, covers a selection of 5,315 journals from both the *Science Citation Index* and *Social Science Citation Index*. (The equivalent in the *SCI-Expanded* would be 7,323 journals.)[2] Furthermore, the year 2003 is the last one available in this database. However, combining some of the tables from the report enables us to construct the following figure:

[2] The *Science Citation Index* covered 5,714 journals in 2003, and the *Social Science Citation Index* 1,708. A subset of 99 journals is covered by both databases.

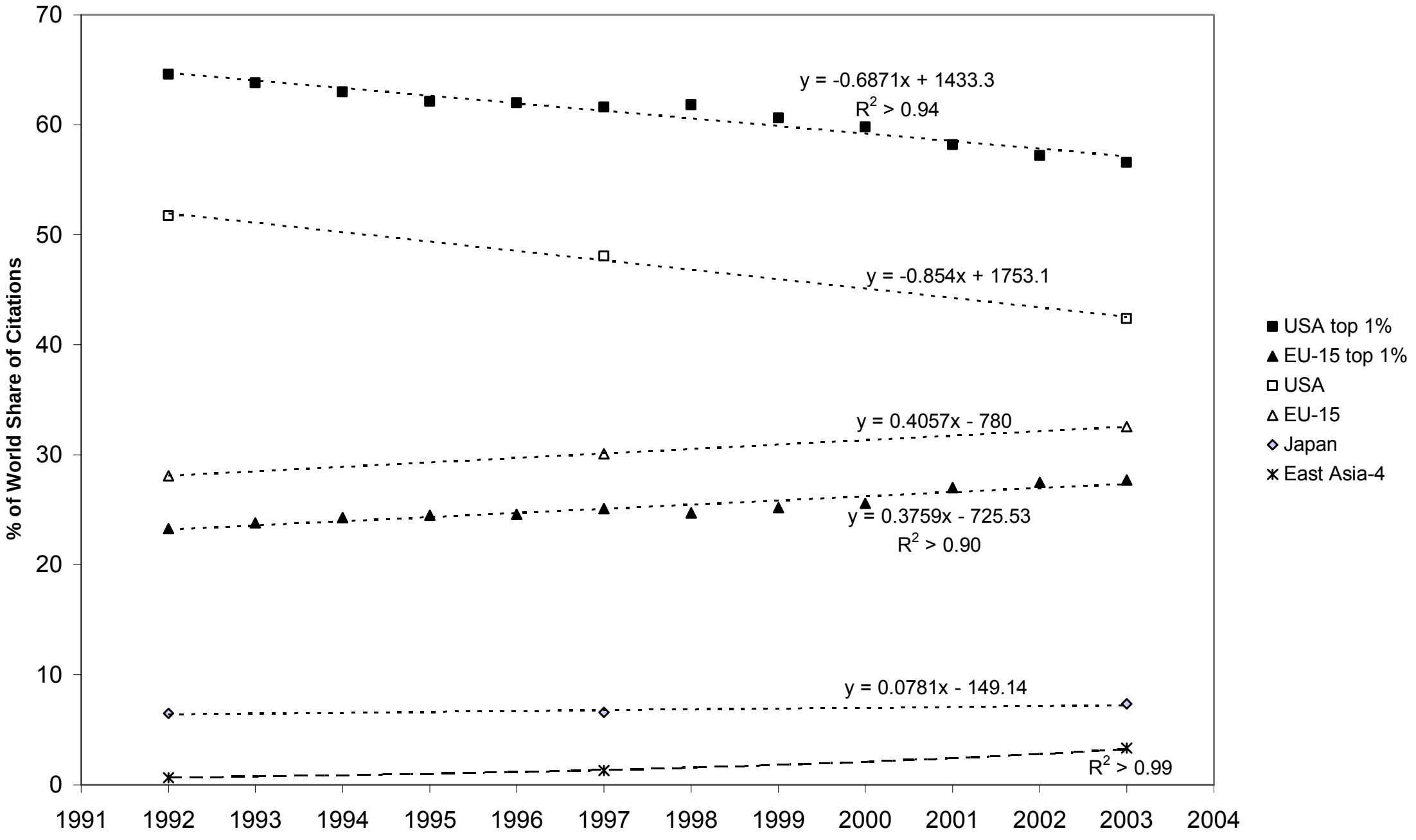

**Figure 6:** Development of the percentage of world shares of citations of the USA, the EU-15, Japan, and the East Asia-4 (China, South Korea, Singapore, and Taiwan). *Source*: NSB (2006, Tables 5-61 and 5-63).

Citation data are available in this report only for 1992, 1997, and 2003, but for the top 1% of most highly cited papers, the in-between years are also provided. The curves show how the values for the EU-15 and the USA have grown together during the decade. The curves for the percentage of world share of 1% most highly cited papers are farther apart than the citation curves, and the slopes are reduced. Thus, the top segment is less affected by these changes than the average ones.

In a first approximation, one could assume that a large part of the 0.4% increase/year of the EU-15 is due to the expansion of the database with mainly European papers, but part of the decrease on the American side would still remain unexplained. (The percentage of world shares of journals with European and American addresses has been very stable with 46.3% for the EU-25 and 38.4% for the USA, respectively.) Thus, the USA is losing ground in terms of number of citations: the average American paper is increasingly similar to the average European one in terms of being cited. Note that the values for the East Asia-4 (China, South Korea, Singapore, and Taiwan) fit an exponential curve again with an $r^2$ larger than 0.99.

## 4. Nanoscience and nanotechnology

The delineation of an emerging field like nanoscience and nanotechnology in terms of a relevant journal set is not a *sine cure* given the interdisciplinarity (between chemistry and applied physics) of this subject area (Braun *et al.*, 1997; Meyer & Persson, 1998; Zitt & Bassecoulard, 2006; Mohrman & Wagner 2006; Braun *et al*., 2007; Mogoutov & Kahane, 2007; Porter *et al*., in preparation). The U.S. Patent and Trade Office (USPTO) decided in 2004 to introduce a new category (Class 977) into its classification scheme devoted to "nanotechnology." Patents issued before this date have been reclassified. Similar efforts have been under way in the European and Japanese Patent Offices, and in international classification schemes (Scheu *et al*., 2006).

In another context (Leydesdorff & Zhou, 2007), one of us analyzed this patent data in more detail. As could be expected, American inventors and assignees are overrepresented in the USPTO-database, while European ones are similarly overrepresented in the EPO-database (Criscuolo, 2006). However, the USPTO database can also be considered as a window on the remainder of the world (Granstrand, 1999; Jaffe & Trajtenberg, 2002). From this perspective, the virtual absence of European patent holders in the "nano" category of this database is remarkable. Among the 1,027 (co-)inventors of the 336 patents classified as "nano" in 2005, 152 came from Japan, but only 33 from Germany.[3] Other Asian nations are represented to a larger extent than European nations. A similar, but even more pronounced pattern can be made visible for the national distribution of the assignees. China is less active in patenting than Japan, Taiwan or South Korea. In this paper, however, we focus on scientific publications (cf. Leydesdorff, 2007).

Using "betweenness centrality" as a measure in the relevant networks of aggregated journal-journal citations (Freeman, 1977; Leydesdorff, forthcoming), we are able to follow the major players in the field of nanoscience as represented in the set of ten core journals (see Table 1 above ). Using the same limitations on document types, Figure 7 can be constructed. We limited the period to 2002-2006 because major initiatives for establishing priority programs in this area were launched in 2000 and 2001. For example, under President Bill Clinton, the U.S. Government launched an initiative in 2000 to promote nanotechnology entitled the *National Nanotechnology Initiative: Leading to the Next Industrial Revolution.* The EU countries, China, Japan, and South Korea, have all adopted nanotechnology as an S&T policy priority. The Chinese government declared nanotechnology a critical R&D priority in their *Guidance for National Development* in 2001. The launch of new journals followed upon the increased funding (Leydesdorff *et al.*, 1994).

[3] In 2005, the database of EPO published 113 patents in the new IPC-class for nanotechnology (B82), and the database of the WIPO 124 patents. The total worldwide was 1,539 patents with publication date in 2005. Class B82 can be considered equivalent to Class 977 in the USPTO database. The EPO recently developed the code "Y10N" as an additional tag to the existing database for the nano-categories (Scheu *et al.*, 2006; Hullmann, 2006).Using this tag, 9,671 patents can be retrieved with publication dates in 2005 (of which 1,196 in the WIPO database and 988 in the EPO one).

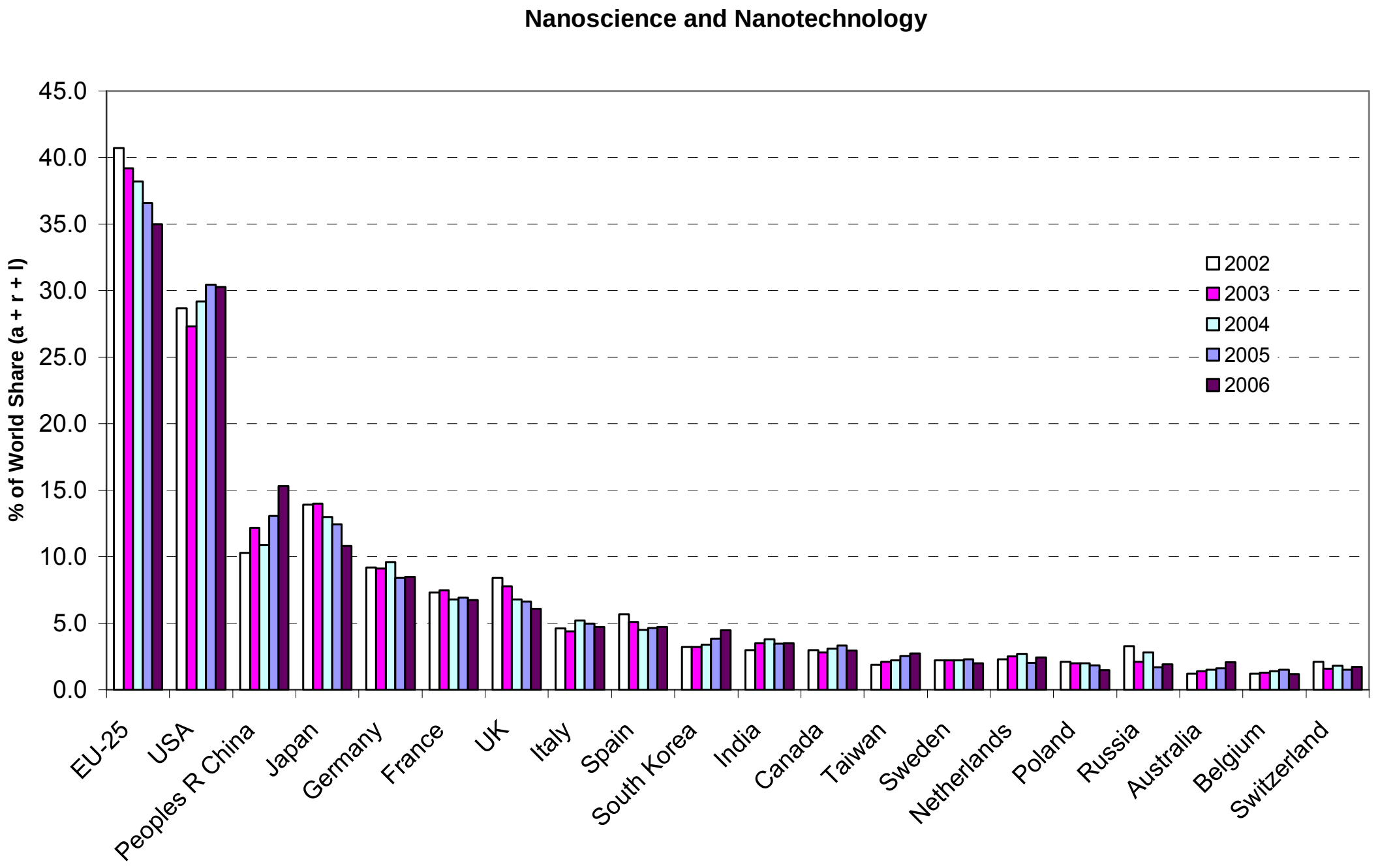

**Figure 7:** Percentage of world share of publications for leading countries and the EU-25 in nanoscience, 2002-2006.

Within this set, the EU-25 is losing more than one percent of its world share of publications per year. The percentage of contributions with an American address has increased since 2003. Similarly, the PRC is gaining ground at the expense of Japan and the leading European nations (including Russia) (Mohrman & Wagner 2006).

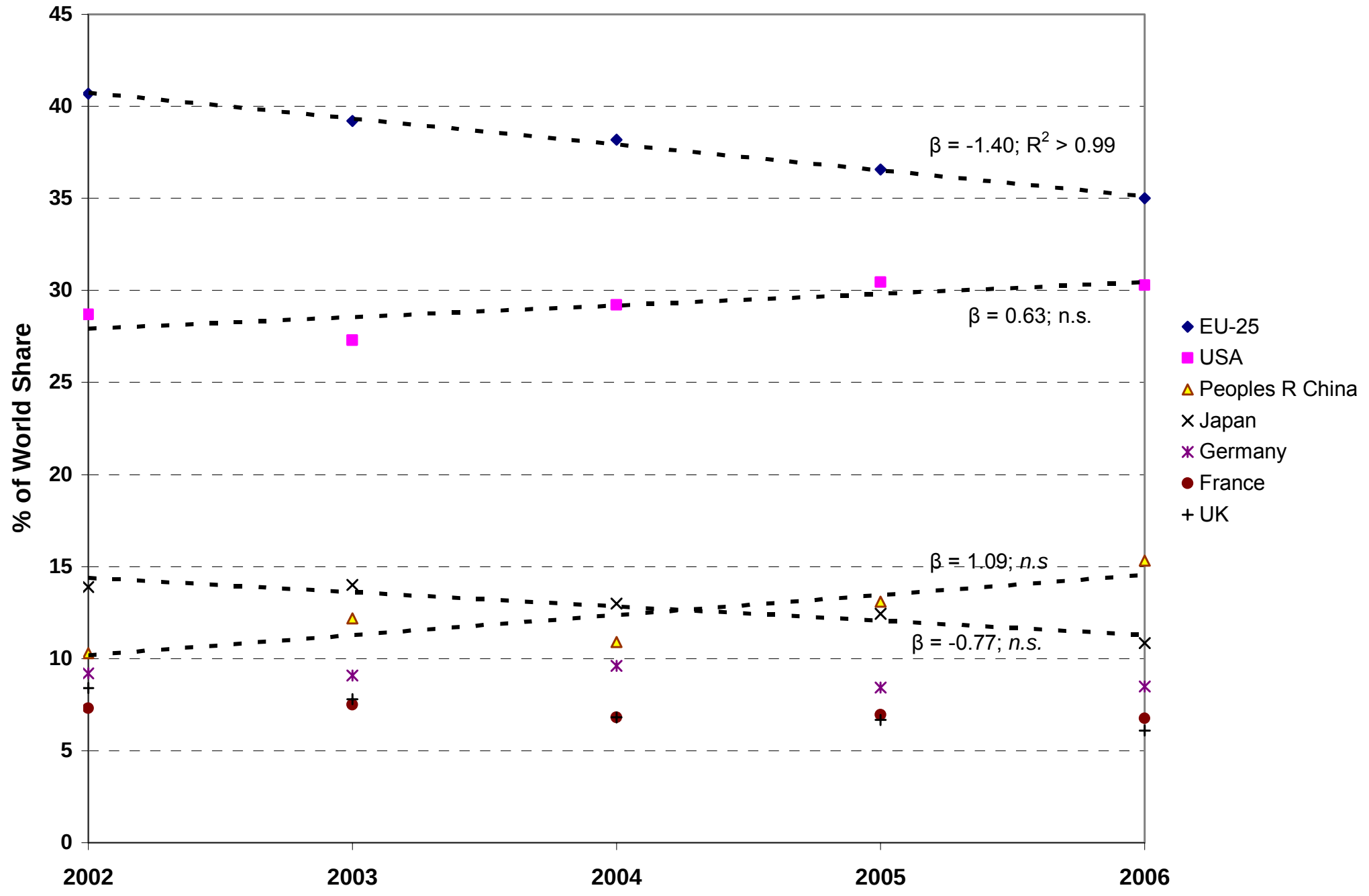

**Figure 8**: Development of the percentage of world share of publications in the domain of ten major "nano"-journals.

In 2006, the percentage of world share for China within this set is 15.3, significantly outperforming Japan (10.8%). However, the growth of China in this domain is not exponential, and the US growth is at least as strong in absolute numbers. The real worry is the decline in the contribution of the EU nations.  It seems to conform what one calls "the European paradox:" the EU is less able to use its research potentials strategically when compared with the USA (Dosi *et al*., 2006).

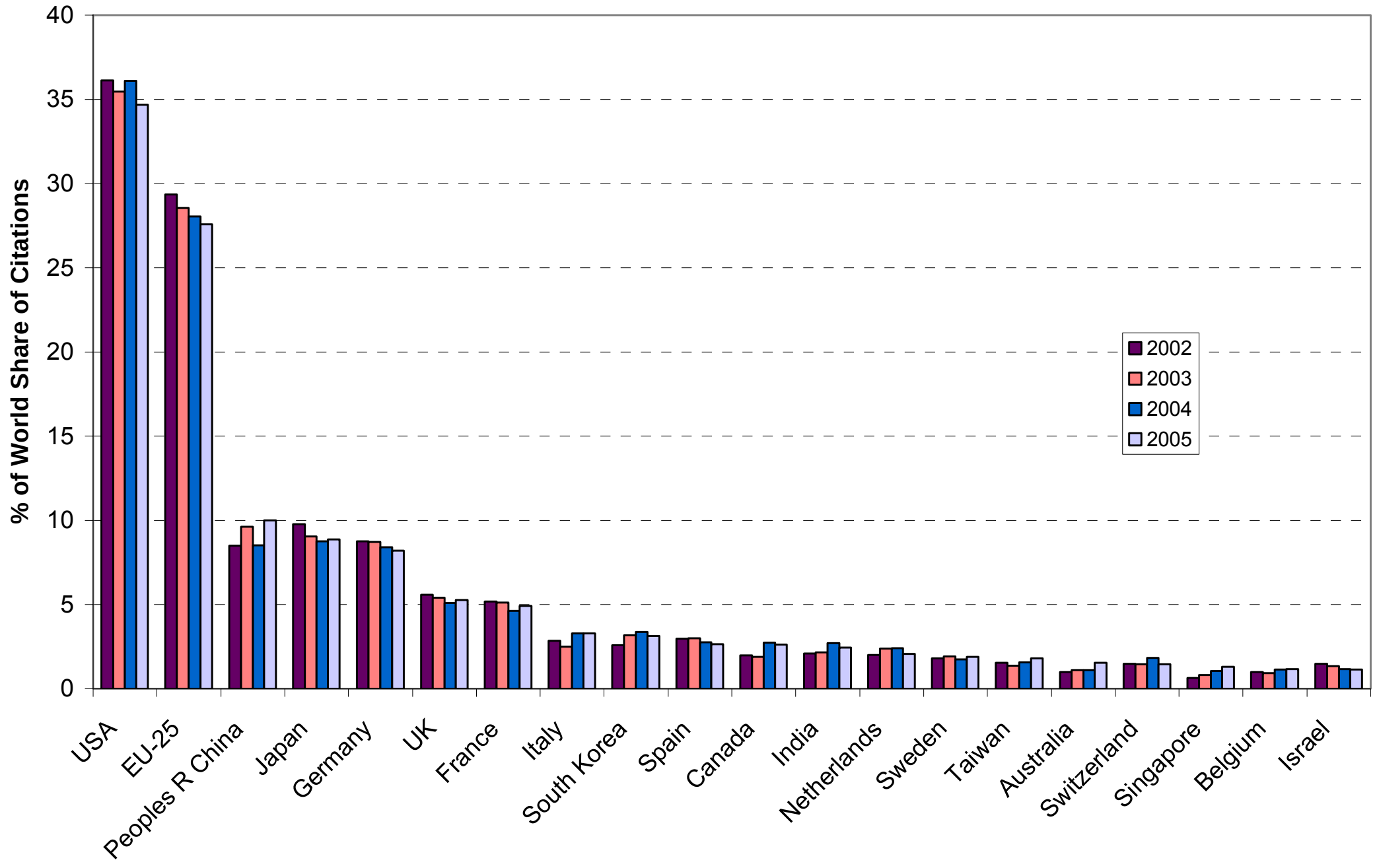

**Figure 9**: Percentage of world share of citations for leading countries and the EU-25 in the set of ten core "nano" journals, 2002-2004.

Figure 9 is based on measuring online the citation rates of all articles, reviews, and letters published in these ten journals during the years 2002-2005.[4] All measurement was done on October 1, 2006. The figure shows that in terms of citations the order between the EU-25 and the USA is reversed with respect to the number of publications. Unlike the general trend (Figure 6), the EU-25 is not improving its citation performance in this specialty area, but rather losing ground more rapidly than the USA to the East Asia-4.

[4] The numbers are 5,807, 6,215, 6,788, and 9,013 documents for the four respective years. These 27,829 documents were cited 280,755 times in total until October 1, 2006. These results should be considered as statistical because the citation counts in the field "times cited" are machine-generated. However, one can expect that the error thus induced is auto-correlated for consecutive years and therefore less affecting the general trend (Bornmann *et al*., forthcoming).

## 5. Conclusions

The contribution of United States scientific institutions to refereed publications indexed by the Institute for Scientific Information has continued to rise in sheer numbers over the past decade, as have most countries. At the same time that the USA and other scientifically advanced countries have maintained slow growth, some countries that are newly developing their own scientific systems are making spectacular gains in numbers of publications contributed to refereed journals. As this happens, the USA and the EU drop as a percentage share of all publications. The drop in percentage share is not an absolute loss of ground.

In terms of citations, papers from the USA and the EU are becoming more equal on average, but the two major players are still far apart in the top segment of the 1% most highly cited papers. The USA is much more successful than the EU in coordinating its research efforts in strategic priority areas like nanotechnology. On all indicators, that is, absolute and relative, China shows exponential growth, while South Korea, Singapore, and Taiwan follow with sustained (mostly linear) growth during the past decade. In nanotechnology, China now ranks as the second nation behind the USA both in terms of number of publications and in its world share of citations.

Our conclusions accord with the conclusions of Braun & Dióspatoni (2005) that in terms of gatekeepers like editorial positions the dominance of the USA is unchallenged (Braun *et al*., 2007). However, the data shows that the scientific system as a whole is growing, and new members are contributing to the pool of knowledge. As they do, the system as a whole benefits. Science is codified and networked at the global level, so it would be difficult to argue that any nationally defined contribution can "lose" in relation to any other part through the addition of new knowledge (Wagner & Leydesdorff, 2005). Far from losing ground in science to new entrants, the USA and other scientifically-advanced countries are gaining new colleagues and partners as well as access to new resources as other countries develop their scientific capacities.